\documentclass[11pt]{article}

\newcommand{\fw}[1]{\textcolor{black}{#1}} 
\newcommand{\kw}[1]{\textcolor{black}{#1}} 
\newcommand{\rev}{\textcolor{black}}

\usepackage[margin=0.5in]{geometry}
\usepackage[figuresright]{rotating}
\usepackage{color}
\usepackage{bm}
\usepackage{amsmath}
\usepackage{graphicx}
\usepackage{setspace}
\usepackage{xr}
\usepackage{hyperref} 
\usepackage{natbib}

\title{Bayesian sample size determination for diagnostic accuracy studies}

\author{Kevin J. Wilson$^{1}$\footnote{email: kevin.wilson@ncl.ac.uk}, S. Faye Williamson$^{2}$, A. Joy Allen$^{3,4}$, \\ 
	Cameron J. Williams$^{1,3,4}$, Thomas P. Hellyer$^{4}$, B. Clare Lendrem$^{3,4}$} 

\date{$^{1}$School of Mathematics, Statistics and Physics, Newcastle University, U.K. \\
	$^{2}$Biostatistics Research Group, Population Health Sciences Institute, Newcastle University, U.K. \\
	$^{3}$NIHR Newcastle In Vitro Diagnostics Co-operative, Newcastle University, U.K. \\ 
	$^{4}$Translational and Clinical Research Institute, Newcastle University, U.K.}

\begin{document}
	
\maketitle

\abstract{The development of a new diagnostic test ideally follows a sequence of stages which, amongst other aims, evaluate technical performance. This includes an analytical validity study, a diagnostic accuracy study and an interventional clinical utility study. \fw{In this paper}, we propose a \fw{novel} Bayesian approach to sample size determination for the diagnostic accuracy study, which takes advantage of information available from the analytical validity stage. We utilise assurance to calculate the required sample size based on the target width of a posterior probability interval and can choose to use or disregard the data from the analytical validity study when subsequently inferring \fw{measures of test accuracy}. \fw{Sensitivity analyses are performed to assess} the robustness of the proposed sample size to the choice of prior, and \fw{prior-data conflict is evaluated by comparing the data to the prior predictive distributions}. We illustrate the \fw{proposed} approach \fw{using a motivating real-life} application involving a diagnostic test for ventilator associated pneumonia. Finally, we compare the properties of the approach against commonly used alternatives. \fw{The results show that,} \kw{when suitable prior information is available}, the assurance-based approach \kw{can reduce the required sample size} when compared to alternative approaches.}

{\bfseries Keywords:} Bayesian assurance, Binomial intervals, Contingency tables, Power calculations, Sensitivity, Specificity.

\section{Introduction}
\label{intro}

Diagnostic accuracy studies evaluate the ability of a diagnostic test (the index test) to correctly identify patients with and without a target condition. This is typically achieved by prospectively comparing results from the index test to the true disease status obtained from the best available reference standard for a cohort of patients. The two main measures used to assess intrinsic diagnostic accuracy are sensitivity and specificity. For a test to proceed to the next stage of evidence development, it is important that these measures are estimated to an appropriate degree of accuracy. This hinges on the sample size chosen for the diagnostic accuracy study. 
Too small a sample size will lead to an imprecise estimate with wide corresponding intervals, which is non-informative to the decision maker, and contributes to research waste \citep{Ioannidis2014}. Alternatively, too large a sample size may delay the results of the study due to longer recruitment times and resource limitations, in addition to financial and ethical implications \citep{Altman1980}. Consequently, choosing a sample size which strikes a balance between accuracy and efficiency is a crucial step in the design of any diagnostic accuracy study. 

Traditional sample size calculations are based on a hypothesis-testing framework. 
The idea is to choose a sample size such that the probability of rejecting the null hypothesis when there is a clinically relevant difference is greater than a required power (typically 80\% or 90\%) with a specified type I error rate (typically 5\% for a two-sided test) \citep{Kunzmann2021}. However, a sample size which captures the precision of the measure of interest, by targeting a desirable width of the corresponding confidence interval, \kw{can be more appropriate in certain circumstances} \citep{Jiroutek2003,Jia2015}. This is pertinent in early clinical diagnostic studies, where the aim is to estimate test accuracy with sufficient precision, which is the approach adopted here. 

In this paper, we consider the sample size problem from a Bayesian perspective and propose a novel approach, referred to as the \textit{Bayesian Assurance Method (BAM)}, to determine sample sizes for diagnostic accuracy studies. In doing so, we explore whether utilising information from the preceding laboratory study will reduce the sample size in the diagnostic accuracy study, and thus lead to a more efficient development process. This may be important if there is need to deploy accurate diagnostic tests rapidly, such as in response to the COVID-19 pandemic, where early detection of infectious individuals is critical to outbreak containment \citep{Rosenthal2020}. 
Another relevant area is rare diseases, where there is a limited number of patients available, or where there are practical or ethical issues with conducting large studies. This extends to (rare) disease subgroups, in which the sensitivity and specificity of a diagnostic test can vary \citep{Bachmann2006}. 

\kw{The BAM shares similar characteristics to seamless and adaptive designs, in that it utilises data from one stage to inform decisions in the subsequent stages in order to improve efficiency and flexibility.} Seamless designs, which aim to combine separate studies, and adaptive designs, which allow for prespecified modifications to the design based on accruing data, are well-established in interventional studies, yet have received little attention in the context of diagnostics. However, the flexibility offered by these designs is just as important in diagnostic accuracy studies. Motivated by the desire to accelerate diagnostic research, \cite{Vach2021} and \cite{Zapf2020} discuss the utility of seamless and adaptive designs, respectively, in developing diagnostics. \cite{Zapf2020} advocate the development and implementation of adaptive designs for diagnostics, and highlight this as a promising area for future research, \kw{which this paper contributes towards.}

The BAM can be used to choose the sample size according to both sensitivity and specificity criteria simultaneously, rather than separately as in most existing methods. Criteria for combining sensitivity and specificity to define the success of a diagnostic test, and how this affects the sample size required, are discussed by \cite{Vach2012}. Korevaar \textit{et al} (2019) suggest specifying a joint hypothesis on the sensitivity and specificity based on predefined minimally acceptable criteria \citep{Korevaar2019}.\cite{Bra07} proposed an approach to choose the sample size based on the predictive probability that the  posterior probability of the sensitivity and specificity both being within pre-specified limits is high. \kw{Although the assurance approach in this paper is related to that taken by \cite{Bra07}, there are some key differences. For example, they required the estimated sensitivity and specificity, along with the upper and lower limits for both intervals, to be specified in advance, and focused only on a two-sided approach, whereas we assure the widths of the intervals directly, requiring only the prior distributions for the parameters, and consider both the one- and two-sided cases.}

Several existing approaches consider binomial confidence intervals based on a normal approximation to determine the sample size (referred to as the Wald interval)\citep{Zhou2011} or some adjustment to it, e.g.\ the Agresti-Coull interval\citep{Agr98}. 
An alternative is to use an exact binomial interval (known as the 
Clopper-Pearson interval \citep{Newcombe2012}). 
A description of commonly used intervals for proportions is provided in \cite[Chapter~3]{Newcombe2012}. \cite[Chapter~4]{Zhou2011} recommend the \cite{Zhou2008} interval for values of sensitivity or specificity close to zero or one. Another recommended interval is the equal-tailed Jeffreys interval\citep{Brown2001}, constructed using a Bayesian approach with a non-informative Jeffreys prior (i.e., Beta(1/2,1/2)) for the binomial proportion. \cite{WeiHutson2013} provide a sample size calculation based on the conditional expectation of interval width given a hypothesised proportion. We compare the BAM to some of these approaches in Section~\ref{comparison}.


Sample size determination from a Bayesian perspective is typically based on \textit{assurance}, which is considered an alternative to power \citep{Oha01}. Assurance, and modifications to it, can be referred to as the probability of success \citep{Bertsche2019}
and the expected/average power \citep{Kowalski2019}, amongst others; a review is provided in \cite[Section~5]{Kunzmann2021}. Unlike power, which is conditional on the true (but unknown) parameter value, the distinguishing property of assurance is that it is an \textit{unconditional} probability which incorporates parameter uncertainty through a prior distribution and integration over the parameter range \citep{Chen2017}. This is formally defined in Section~\ref{assurance_defn}. 

The use of assurance for sample size calculations has occurred predominantly within clinical trials \citep{Kunzmann2021}. In this paper, we use assurance to represent the probability of obtaining the desired accuracy (based on a target interval width) in our estimates of sensitivity and/or specificity. The sample size is then taken to be the minimum which yields the required assurance. 
We describe inference for a standard diagnostic accuracy study in Section \ref{inference}. The BAM is presented and further described in Section~\ref{assurance_defn}, with issues such as prior sensitivity and prior-data conflict addressed in Section~\ref{prior_sensitivity}. As a motivating case study, we use the BAM to redesign a diagnostic accuracy study of a test for ventilator associated pneumonia in Section~\ref{application}, and assess the properties of the BAM, in comparison to some standard approaches, in Section \ref{comparison}. 

\section{Inference in a diagnostic accuracy study}
\label{inference}

We consider a diagnostic accuracy study to assess an index test under development. In the study, we observe the numbers of individuals in a $2\times2$ contingency table (Table \ref{2by2} (i)).

\begin{table}[ht]
	\begin{center}
		\resizebox{0.4\textwidth}{!} {
			\centering
			\begin{tabular}{|l|cc|c|} \hline
				(i) & Disease & No Disease & Total \\ \hline
				Test Positive & $n_{1,1}$ & $n_{1,2}$ & $n_{1,T}$ \\
				Test Negative & $n_{2,1}$ & $n_{2,2}$ & $n_{2,T}$ \\ 
				Total & $n_{T,1}$ & $n_{T,2}$ & $n_T$ \\ \hline 
				(ii) &  VAP & No VAP & Total \\\hline
				Test Positive & 16 & 35 & 51 \\
				Test Negative & 1 & 20 & 21 \\ 
				Total  & 17 & 55 & 72 \\ \hline 
				(iii) & VAP & No VAP & Total \\ \hline
				Test Positive & 51 & 55 & 106 \\
				Test Negative & 2 & 42 & 44 \\ 
				Total  & 53 & 97 & 150 \\ \hline 
			\end{tabular}
		}
		\caption{(i) A $2\times2$ contingency table for a typical diagnostic accuracy study.
			(ii) The $2\times2$ contingency table for the {\em biomarker selection study} based on the biomarker IL-1$\beta$.
			(iii) The $2\times2$ contingency table for the {\em diagnostic accuracy study} based on the biomarker IL-1$\beta$.}
		\label{2by2}
	\end{center}
\end{table}

The number of individuals with and without the disease is assumed to be known, based on a reference test. The intrinsic accuracy of the index test can be measured by its sensitivity and specificity, \fw{defined as} the probability of a positive test given disease and the probability of a negative test given no disease, respectively.

There are two approaches used to model numbers of individuals in the cells of the $2\times2$ table:\ assuming either binomial or multinomial likelihoods. In the first case,
$
n_{1,1}\mid\lambda,n_{T,1}  \sim  \textrm{Bin} (n_{T,1},\lambda) \text{ and } 
n_{2,2}\mid\theta,n_{T,2}  \sim  \textrm{Bin} (n_{T,2},\theta),
$
where $\lambda$ is the sensitivity and $\theta$ is the specificity of the index test. The conjugate prior distributions are $\lambda\sim\textrm{Beta}(a_\lambda,b_\lambda)$ and $\theta\sim\textrm{Beta}(a_\theta,b_\theta)$. If we assume in the prior that the sensitivity and specificity are independent, then their posterior distributions are
$\lambda\mid\bm n  \sim  \textrm{Beta}(a_\lambda+n_{1,1},b_\lambda+n_{2,1})$  and  
$\theta\mid\bm n  \sim  \textrm{Beta}(a_\theta+n_{2,2},b_\theta+n_{1,2})$.
The independence assumption will often be reasonable since the diagnostic thresholds for the test are fixed at this stage, and the sensitivity and specificity consider mutually exclusive populations of patients.

In the second case, we consider the vector $\bm n = (n_{1,1},n_{1,2},n_{2,1},n_{2,2})^{'}$ and assume
$
\bm n\mid\bm\gamma \sim \textrm{Multi}(n,\bm\gamma),
$
where $\bm\gamma=(\gamma_{1,1},\gamma_{1,2},\gamma_{2,1},\gamma_{2,2})^{'}$ is a vector containing the probabilities of each cell of the contingency table. \fw{Here,} the sensitivity and specificity are given by
$ \lambda = \gamma_{1,1} / ( \gamma_{1,1} + \gamma_{2,1} ) $ and 
$ \theta = \gamma_{2,2} / ( \gamma_{1,2} + \gamma_{2,2} ) $.
A typical form for the prior distribution is a Dirichlet distribution, which provides conjugacy. That is, $\bm\gamma\sim\textrm{Dir}(\bm\alpha)$, where $\bm\alpha=(\alpha_{1,1},\alpha_{1,2},\alpha_{2,1},\alpha_{2,2})^{'}$. 

It can be shown that the two approaches are equivalent in terms of inference for the sensitivity and specificity (see Appendix A). \fw{In this paper, we will} use the binomial form as it allows for the direct specification of the priors for the sensitivity, specificity and prevalence. \rev{We will assume conjugate beta priors, as detailed above, throughout the rest of the paper.}

\section{Sample size determination}
\label{assurance_defn}

\subsection{Assurance} 

Assurance \fw{is} a Bayesian alternative to power to choose a sample size. 
Consider a two-armed clinical trial in which a hypothesis test is to be conducted with $\textrm{H}_0:\rev{\delta}=0$ versus $\textrm{H}_1:\rev{\delta}>0$, \fw{where} \rev{$\delta$} represents the difference in the effect of two treatments. A typical power calculation would choose a sample size to provide a certain statistical power at a particular assumed value \rev{$\delta_c$} for \rev{$\delta$}, often taken to be the minimal clinically relevant difference. In this case, the power is \rev{$\Pr(\textrm{Reject H}_0\mid\delta=\delta_c)$} and would increase with sample size.

In practice, the choice of \rev{$\delta_c$} is relatively arbitrary. As the true effect size $\delta$ is unknown, this can result in conditioning on an event which is extremely unlikely. \rev{One approach to mitigate this is to conduct a sensitivity analysis, varying the value of $\delta_c$ and choosing a sample size which is robust to small perturbations\citep{Mat06}.} In the Bayesian context we can \rev{take an alternative approach, and} represent our uncertainty over $\delta$ using a prior distribution $\pi(\delta)$. The assurance is the expected power of the hypothesis test with respect to this prior,
$
A(n)  =  \textrm{E}_{\delta}\left[\Pr(\textrm{Reject H}_0\mid \delta)\right] =  \int \Pr(\textrm{Reject H}_0\mid \delta)\pi(\delta) d\delta.
$
We choose to make the dependence on the sample size $n$ explicit for the assurance $A(\cdot)$.

Assurance is not restricted to where we will perform a hypothesis test at the end of a trial. If we perform a Bayesian analysis instead, then we may declare the trial a success and the new treatment superior if $\Pr(\delta\leq0)\leq0.05$ in the posterior, for example. In this case,
$
A(n)  =  \textrm{E}_{\delta}\left[\Pr(\textrm{Trial a success}\mid \delta)\right] =  \int \Pr(\textrm{Trial a success}\mid \delta)\pi(\delta) d\delta.
$
Thus, the assurance is the unconditional probability that the trial results in a successful outcome.

We use assurance to choose a sample size to estimate sensitivity, specificity, or both, of the index test to a certain degree of accuracy. 
We initially focus on \textit{sensitivity} of the index test, $\lambda$ and consider two cases:\ assuring the width of the posterior probability interval (two-sided), and assuring the width of the \textit{lower half} of the posterior probability interval (one-sided).

\subsubsection{Two-sided \fw{case}}
Considering the inference from Section \ref{inference}, a $100(1-\alpha)\%$ symmetric posterior probability interval for $\lambda$ is $(\lambda_L,\lambda_U)$, where the limits of the interval are defined such that
$
\Pr(\lambda\leq\lambda_L\mid \bm n)  =  \dfrac{\alpha}{2} \text{ and }
\Pr(\lambda\geq\lambda_U\mid \bm n)  =  \dfrac{\alpha}{2}.
$
The accuracy of the estimation of $\lambda$ can be considered as the width of this interval, $\lambda_U-\lambda_L$, and a successful diagnostic accuracy study would produce an interval with a width smaller than some target, $\lambda_U-\lambda_L\leq w^{*}$.

Suppose the number of individuals with the disease in the study, $n_{T,1}$, is fixed. There are three possibilities: no values of $n_{1,1}$ lead to an interval with width smaller than $w^{*}$, all values of $n_{1,1}$ lead to an interval with width smaller than $w^{*}$, or some values of $n_{1,1}$ lead to an interval with width smaller than $w^{*}$. To investigate the third case, consider the posterior variance of $\lambda$,
$\textrm{Var}(\lambda\mid \bm n) = \{(a_\lambda+n_{1,1})(b_\lambda+n_{2,1})\}/\{(a_\lambda+b_\lambda+n_{T,1})^2(a_\lambda+b_\lambda+n_{T,1}+1)\}.$
For a fixed sample size $n_{T,1}$, the denominator of this fraction is constant. That is,
$
\textrm{Var}(\lambda\mid \bm n)  \propto (a_\lambda+n_{1,1})(b_\lambda+n_{2,1}) \propto n_{1,1}(b_\lambda-a_\lambda+n_{T,1})-n_{1,1}^2,
$
substituting $n_{2,1}=n_{T,1}-n_{1,1}$. The variance is quadratic in $n_{1,1}$ and the squared term has a negative coefficient. 
Thus, the posterior probability interval will be narrower than $w^{*}$ when $n_{1,1}\leq c_1$ and $n_{1,1}\geq c_2$, for two critical numbers of individuals $c_1<c_2$. We define this set as $\mathcal{N}=\left\{n_{1,1}: n_{1,1}\leq c_1 \textrm{ or } n_{1,1}\geq c_2\right\}$.

\subsubsection{One-sided \fw{case}}

We consider a $100(1-\alpha)\%$ posterior probability interval for $\lambda$ of the form $(\lambda_L,1)$, where the lower limit of the interval is defined such that $\Pr(\lambda\leq\lambda_L\mid \bm n)  = \alpha$.
We consider the distance between the lower limit of the interval and a central point estimate of $\lambda$, i.e.\ $\lambda_{0.5}-\lambda_L$, where $\lambda_{0.5}$ is the posterior median. A successful diagnostic accuracy study would result in this interval having a width smaller than some target, $\lambda_{0.5}-\lambda_L\leq w^{*}$.

By the same logic as the \fw{two-sided} case, the posterior probability interval will be narrower than $w^{*}$ when $n_{1,1}\leq c_1$ and $n_{1,1}\geq c_2$, for two critical numbers of individuals $c_1<c_2$. \fw{Thus, we consider the set $\mathcal{N}=\left\{n_{1,1}: n_{1,1}\leq c_1 \textrm{ or } n_{1,1}\geq c_2\right\}$ for the one-sided case, with $c_1$ and $c_2$ determined by the interval $\lambda_{0.5}-\lambda_L$.}

\subsubsection{Evaluating the assurance} \label{evaluating}

We can \fw{obtain} an expression for the assurance for a sample size $n_T$, conditional on a fixed number of diseased individuals $n_{T,1}$. \fw{This is denoted by $A_\lambda(n_{T}\mid n_{T,1})$ and defined as}
\begin{eqnarray}
A_\lambda(n_{T}\mid n_{T,1}) & = & \int \Pr(\textrm{Accuracy achieved}\mid\lambda)\pi(\lambda) d\lambda, \nonumber \\
& = & \dfrac{\Gamma(a_\lambda+b_\lambda)}{\Gamma(a_\lambda)\Gamma(b_\lambda)}\sum_{n_{1,1}\in \mathcal{N}}{n_{T,1} \choose n_{1,1}} \dfrac{\Gamma(a_\lambda+n_{1,1})\Gamma(b_\lambda+n_{2,1})}{\Gamma(a_\lambda+b_\lambda+n_{T,1})}, \label{ass.sens}
\end{eqnarray}
where $\Gamma(\cdot)$ represents the gamma function. A derivation is given in Section A of the supplementary material.

As the number of individuals with the disease, $n_{T,1}$, will not be known in advance, we need to sum over the possible values $n_{T,1}$ can take. If we have a random sample from the target population, then $n_{T,1}\mid\rho\sim\textrm{Bin}(n_T,\rho)$,
where $\rho$ is the prevalence of the disease. Let $\rho \sim \textrm{Beta}(a_\rho,b_\rho)$ for some chosen values of $(a_\rho,b_\rho)$. The unconditional assurance is then
\begin{eqnarray*}
A_\lambda(n_{T}) & = & \sum_{n_{T,1}=0}^{n_T}\left[\int \Pr(\textrm{Accuracy achieved}\mid\lambda)\pi(\lambda)d\lambda\times f(n_{T,1})\right],
\end{eqnarray*}
where $f(n_{T,1})=\int f(n_{T,1}\mid\rho)\pi(\rho)d\rho$ is the probability of observing $n_{T,1}$ individuals in the disease group. The assurance can \fw{thus} be expressed as
\begin{multline}
A_\lambda(n_{T}) = \dfrac{\Gamma(a_\lambda+b_\lambda)}{\Gamma(a_\lambda)\Gamma(b_\lambda)}\sum_{n_{T,1}=0}^{n_T}\bigg\{ \sum_{n_{1,1}\in \mathcal{N}}\left[{n_{T,1} \choose n_{1,1}} \dfrac{\Gamma(a_\lambda+n_{1,1})\Gamma(b_\lambda+n_{2,1})}{\Gamma(a_\lambda+b_\lambda+n_{T,1})}\right] \\
\times {n_T \choose n_{T,1}}\dfrac{\Gamma(a_\rho+b_\rho)}{\Gamma(a_\rho)\Gamma(b_\rho)}\dfrac{\Gamma(a_\rho+n_{T,1})\Gamma(b_\rho+n_{T,2})}{\Gamma(a_\rho+b_\rho+n_T)} \bigg\}. 
\label{asssens}
\end{multline}
This is derived in Section A of the supplementary material.

All that remains is to find the values of $(c_1,c_2)$. For each fixed sample size, $n_T$, and number of diseased individuals, $n_{T,1}$, the values of \fw{$\lambda_L,\lambda_{0.5}$ and $\lambda_U$} will depend only on $n_{1,1}$ and, hence, the width of the interval will be a function of $n_{1,1}$, $W(n_{1,1})$, in both cases. Therefore, $c_1 =  \textrm{argmin}\left\{W(n_{1,1})\geq w^{*}\right\} - 1$ and $c_2 = \textrm{argmax}\left\{W(n_{1,1})\geq w^{*}\right\} + 1$ for $n_1<n_{T,1}<n_2$, where $n_1$ is a number below which the interval can never achieve the desired width and $n_2$ is a number above which the width of the interval is always below $w^{*}$. Hence, $A_\lambda(n_T\mid n_{T,1})=0$ for all  $n_{T,1}\leq n_1$ and  $A_\lambda(n_T\mid n_{T,1})=1$ for all $n_{T,1}\geq n_2$.

To estimate the \textit{specificity} of the index test to a given accuracy of $w^{*}$, we can derive the assurance in the same way, \fw{which results in an assurance analogous to that in equation \eqref{asssens}}. 
The details are given in Section A of the supplementary material. 

\fw{Finally, suppose} we wish to estimate \textit{both} the sensitivity and specificity to a particular accuracy. 
Consider different accuracy targets, $w_\lambda^{*}$ and $w_\theta^{*}$, for the sensitivity and specificity, respectively. In this case, the assurance for the sample size $n_T$ conditional on $n_{T,1}$ (and hence $n_{T,2}$, since $n_{T,2}=n_T-n_{T,1}$) is given by 
\begin{eqnarray*}
A_{\lambda,\theta}(n_{T}\mid n_{T,1}) & = & \int\int \Pr(\textrm{Accuracy achieved}\mid\lambda,\theta)\pi(\lambda)\pi(\theta) d\lambda d\theta, \\
& = & \dfrac{\Gamma(a_\lambda+b_\lambda)}{\Gamma(a_\lambda)\Gamma(b_\lambda)}\sum_{n_{1,1}\in \mathcal{N}_1}{n_{T,1} \choose n_{1,1}} \dfrac{\Gamma(a_\lambda+n_{1,1})\Gamma(b_\lambda+n_{2,1})}{\Gamma(a_\lambda+b_\lambda+n_{T,1})} z \\ &&
\times\dfrac{\Gamma(a_\theta+b_\theta)}{\Gamma(a_\theta)\Gamma(b_\theta)}\sum_{n_{2,2}\in \mathcal{N}_2}{n_{T,2} \choose n_{2,2}} \dfrac{\Gamma(a_\theta+n_{2,2})\Gamma(b_\theta+n_{1,2})}{\Gamma(a_\theta+b_\theta+n_{T,2})},
\end{eqnarray*}
where $\mathcal{N}_1$ contains the values $n_{1,1}\leq c_{1}$ and $n_{1,1}\geq c_{2}$ that give a posterior interval narrower than $w_\lambda^{*}$ for the sensitivity, and $\mathcal{N}_2$ contains the values $ n_{2,2}\leq \tilde{c}_{1}$ and $n_{2,2}\geq \tilde{c}_{2}$ that give a posterior interval narrower than $w_\theta^{*}$ for the specificity.   

To find the unconditional assurance, we sum over the possible values of $n_{T,1}$ to give:
\begin{multline} 
A_{\lambda,\theta}(n_{T}) = \dfrac{\Gamma(a_\lambda+b_\lambda)}{\Gamma(a_\lambda)\Gamma(b_\lambda)}\dfrac{\Gamma(a_\theta+b_\theta)}{\Gamma(a_\theta)\Gamma(b_\theta)}\sum_{n_{T,1}=0}^{n_T}\Bigg\{\sum_{n_{1,1}\in \mathcal{N}_1}\left[{n_{T,1} \choose n_{1,1}} \dfrac{\Gamma(a_\lambda+n_{1,1})\Gamma(b_\lambda+n_{2,1})}{\Gamma(a_\lambda+b_\lambda+n_{T,1})}\right] \\
\times\sum_{n_{2,2}\in \mathcal{N}_2}\left[{n_{T,2} \choose n_{2,2}} \dfrac{\Gamma(a_\theta+n_{2,2})\Gamma(b_\theta+n_{1,2})}{\Gamma(a_\theta+b_\theta+n_{T,2})}\right]
\times {n_T \choose n_{T,1}}\dfrac{\Gamma(a_\rho+b_\rho)}{\Gamma(a_\rho)\Gamma(b_\rho)}\dfrac{\Gamma(a_\rho+n_{T,1})\Gamma(b_\rho+n_{T,2})}{\Gamma(a_\rho+b_\rho+n_T)}\Bigg\}. 
\label{assboth}
\end{multline}

\fw{The proposed BAM is now summarised via the following steps:}
\begin{enumerate}
    \item Choose whether we wish to assure our estimate of sensitivity $\lambda$, specificity $\theta$, or both.
    \item Choose a \fw{target width(s)} $w^{*}$ for the \fw{accuracy} measure(s), a one- or two-sided posterior interval and a level $\alpha$ for the interval.
    \item Specify the prior distributions for the \fw{chosen accuracy} measure(s) and the prevalence \fw{$\rho$}. We detail how to do this in the next section.
    \item Use \fw{equation} (\ref{asssens}) or (\ref{assboth}) (or see Section A of the supplementary material) to calculate the assurance for sample sizes $n_T=1,2,\ldots$.
    \item Choose the minimum sample size $n_T^{*}$ to give the desired assurance.
\end{enumerate}

\fw{\textit{Example:} Suppose} we wish to estimate both sensitivity and specificity to within $5\%$, with posterior probability $0.99$ using a two-sided interval, i.e.\ $w^{*}=0.05$ and $\alpha=0.01$. We specify prior distributions for $\lambda,\; \theta$ and $\rho$, and use equation (\ref{assboth}) to evaluate the assurance for sample sizes $n_T=1,2,\ldots$. \fw{To achieve the desired accuracy with a probability of at least 0.9, say, we choose the smallest value of $n_T$ which gives rise to an assurance greater than 0.9.}

\section{Prior specification and model checking} \label{prior_sensitivity}


A diagnostic accuracy study \fw{is part of an extensive development process for the diagnostic test \citep[see][Figure 1]{Gra20}}. \fw{Its main purpose} is to estimate performance characteristics of the test, particularly the sensitivity and specificity, in the target population in a clinically relevant setting. Prior to \fw{the diagnostic accuracy study} is the analytic\fw{al} validity phase, \fw{in which} the test may still be under development and the data generated may be used to support regulatory approvals \citep{Gra20}. The validation conducted \fw{during} this stage may \fw{test} individuals from the target population. \fw{Consequently, the data produced can be used to inform the prior distributions in the diagnostic accuracy study. This assumes that the observations in the two stages are exchangeable, which may not always be reasonable. Therefore, in Section B of the supplementary material, we detail how the BAM can be used under weaker assumptions.}

\subsection{Specifying prior distributions}
\label{prior}

Consider the analytical validity testing. Suppose that a random sample of $n_T^0$ individuals was taken and the numbers in the cells of the $2\times2$ \fw{contingency} table were $\bm n^0=(n_{1,1}^0,n_{1,2}^0,n_{2,1}^0,n_{2,2}^0)^{'}$. Using the inferential approach in Section \ref{inference}, priors for the sensitivity, specificity and prevalence would be $\lambda\sim\textrm{Beta}(a_\lambda^0,b_\lambda^0)$, $\theta\sim\textrm{Beta}(a_\theta^0,b_\theta^0)$ and $\rho\sim\textrm{Beta}(a_\rho^0,b_\rho^0)$, \fw{respectively. The corresponding} posterior distributions (excluding conditioning statements) would be
$
\lambda  \sim  \textrm{Beta}(a_\lambda^1,b_\lambda^1), \;
\theta  \sim  \textrm{Beta}(a_\theta^1,b_\theta^1) \text{ and }
\rho  \sim  \textrm{Beta}(a_\rho^1,b_\rho^1), 
\label{beta_priors}
$
where $a_\lambda^1=a_\lambda^0+n_{1,1}^0$, $b_\lambda^1=b_\lambda^0+n_{2,1}^0$, $a_\theta^1=a_\theta^0+n_{2,2}^0$, $b_\theta^1=b_\theta^0+n_{1,2}^0$, $a_\rho^1=a_\rho^0+n_{T,1}^0$ and
$b_\rho^1=b_\rho^0+n_{T,2}^0$. 

\fw{These latter} beta distributions can be used as priors for the diagnostic accuracy study. \fw{Although} this does not negate the necessity of choosing the initial prior values $(a_\lambda^0,b_\lambda^0)$, $(a_\theta^0,b_\theta^0)$ and $(a_\rho^0,b_\rho^0)$, these will have a small effect on the sample size chosen if sufficient data is available from the analytical validity stage. \fw{This is explored further in the next section.} \rev{The approach taken here is equivalent to using a power prior with the parameter quantifying the heterogeneity between the diagnostic study population and analytic validity population set equal to one (representing homogeneous populations). In cases of heterogeneity between the two populations, a power prior could be used with this parameter taking a value in the range $[0,1]$. For full details see \cite{Ibr15}.}

\fw{In cases where it is controversial to use data from the analytical validity stage when inferring the sensitivity and specificity of the test, we could use a weaker prior in the analysis, but retain the original prior in the design to inform the sample size calculations. This is illustrated in Section B of the supplementary material}


\subsection{Prior sensitivity}
\label{sensitivity}
The choice of initial prior parameters, $(a_\lambda^0,b_\lambda^0)$, $(a_\theta^0,b_\theta^0)$ and $(a_\rho^0,b_\rho^0)$ may have little effect on the assurance if sufficient data are observed at the analytic validity stage. We explore this using local sensitivity analysis and \fw{investigate the following two questions:} \\
(1) How does the optimal sample size, $n_T^{*}$, change with values of the prior parameters? \\
(2) How does the assurance at $n_T^{*}$, $A(n_T^{*})$, change with the prior parameters? \\
\fw{In particular,} we vary the prior parameters $(a^0_{C},b^0_{C})$ for $C=\{\lambda, \theta, \rho\}$ in turn over a range of values around their initial values, and record the smallest and largest values of the optimal sample size $(\underline{n_T^{*}},\overline{n_T^{*}})$ and assurance $(\underline{A}(n_T^{*}),\overline{A}(n_T^{*}))$. If these values do not differ by much, then \fw{the optimal sample size is relatively robust to the initial prior choice.}

\fw{Using the grid search approach \citep{Roo15} to determine an appropriate range of prior parameter values,} we explore the sensitivity on a grid $G_{a^0,b^0}(\epsilon)$, where $\epsilon$ represents the distance between a prior and the original prior with parameters $(a^0,b^0)$. That is, 
$G_{a^0,b^0}(\epsilon) = \{(a,b):d(\pi_{a,b}(\gamma),\pi_{a^0,b^0}(\gamma)) = \epsilon\}$,
where $\pi_{a,b}(\gamma)$ represents the beta prior distribution with parameters $(a,b)$ and $\gamma$ is one of $\lambda$, $\theta$ and $\rho$. We use the Hellinger distance \citep{Roo15} which, for the beta distribution, can be expressed as
\begin{displaymath}
d(\pi_{a,b}(\gamma),\pi_{a^0,b^0}(\gamma))=1-\dfrac{B([a+a^0]/2,[b+b^0]/2)}{\sqrt{B(a,b)B(a^0,b^0)}},
\end{displaymath}
where $B(a,b)=\Gamma(a)\Gamma(b)/\Gamma(a+b)$ is the beta function. To conduct the grid search, it is sensible to work in polar co-ordinates. \fw{Therefore}, we set $a=\exp(z)\cos(\phi)$ and $b=\exp(z)\sin(\phi)$, where $z=log(r)$. We search in the range $\phi\in[-\pi,\pi]$, solving for the value of $r$ which gives the correct value of $\epsilon$.

To find the values of $a$ and $b$, we convert back via $a=a^0+r\cos(\phi)$ and $b=b^0+r\sin(\phi)$. From this grid search, we can then find \fw{the corresponding} $(\underline{n_T^{*}},\overline{n_T^{*}})$ and $(\underline{A}(n_T^{*}),\overline{A}(n_T^{*}))$ for this $\epsilon$. We suggest a sensible choice of $\epsilon$ \fw{in Section \ref{prior_sens}.}

\subsection{Prior-data conflict}
\label{conflict}

Label the counts in the $2\times2$ table from the diagnostic accuracy study $\bm n^1 = (n_{1,1}^1,n_{1,2}^1,n_{2,1}^1,n_{2,2}^1)^{'}$. The posterior distributions for the sensitivity and specificity (omitting the conditioning) will be  $\lambda\sim\textrm{Beta}(a_\lambda^2,b_\lambda^2)$ and $\theta\sim\textrm{Beta}(a_\theta^2,b_\theta^2)$, \fw{respectively}, where 
\begin{eqnarray*}
a_\lambda^2 & = a_\lambda^1 + n_{1,1}^1 = a_\lambda^0 +n_{1,1}^0 + n_{1,1}^1, \;\;
b_\lambda^2 = b_\lambda^1 + n_{2,1}^1 = b_\lambda^0 +n_{2,1}^0 + n_{2,1}^1, \\
a_\theta^2 & = a_\theta^1 + n_{2,2}^1 = a_\theta^0 +n_{2,2}^0 + n_{2,2}^1, \;\; 
b_\theta^2 = b_\theta^1 + n_{1,2}^1 = b_\theta^0 +n_{1,2}^0 + n_{1,2}^1.
\end{eqnarray*}

The inference for the sensitivity and specificity is in the form of a weighted average of the prior and the observations, with weights determined by the relative sample sizes of each. The prior is made up of a weighted average of the observations in the analytical validity stage and the original prior. If all of the elements are in broad agreement, then the posterior distribution will provide an accurate summary of the properties of the index test in the population of interest. However, it could be the case that the prior and observations are not in agreement, \fw{which} is known as \textit{prior-data conflict} \rev{\citep{Box80,Sch14}}. For example, if the two studies are carried out \fw{at different times or} in different locations, the spectrum of disease in the target population may not be the same. In this case, it is important to investigate why the differences are there and what action should be taken.

We can evaluate prior-data conflict by comparing the observations to the prior predictive distributions of the parameters. We consider the prior predictive distributions of the number of observations in the disease group, \fw{$n_{T,1}$}, and, conditional on this, the numbers who test positive, $n_{1,1}$ of those with the disease, and the number who test negative of those without the disease, $n_{2,2}$. These are given by $f(y) = {n \choose y}\dfrac{B(a+y,b+n-y)}{B(a,b)}$,
where $y$ is $(\fw{n_{T,1}},n_{1,1},n_{2,2})$ in turn, $n$ is the corresponding sample size, i.e.\ ($n_T,n_{T,1},n_{T,2})$, and $(a,b)$ are the beta distribution parameter values for the prevalence, sensitivity and specificity, respectively. We can then plot the prior predictive distributions and calculate probabilities of the form $\Pr(n\geq n_{\textrm{obs}})$, for observed number of individuals $n_{\textrm{obs}}$. If the observed value lies in the body of the \fw{associated} prior predictive distribution, then that prior is consistent with the data. \fw{Otherwise}, this provides evidence of prior-data conflict.

\section{A biomarker test for ventilator associated pneumonia} \label{application}
Using published results \citep{Mor10,Hel15}, we consider the development of a biomarker test for ventilator associated pneumonia (VAP). The development of the test involved four stages; an exploratory study to look at possible biomarkers for VAP diagnosis
, a single centre observational study to choose suitable biomarkers, 
a multicentre diagnostic accuracy study to develop biomarker cut offs and validate accuracy 
and a randomised controlled trial of clinical utility. 
At each stage the target population was patients on a ventilator with suspected VAP. The reference standard test was the growth of pathogens at $>10^4$ colony forming units per millilitre of bronchoalveolar fluid. All patients with suspected VAP receive antibiotics, although only 20-60\% of patients will have VAP confirmed by the reference standard, leading to overuse of antibiotics. Microbiology culture and sensitivities takes up to 72 hours to return results to clinicians, which delays the opportunity to discontinue antibiotics in patients who do not have infection. A rapid, highly sensitive biomarker test could allow for early stopping of antibiotics.

We consider planning the diagnostic accuracy study. The sample size was originally chosen to reduce the width of the 95\% confidence interval for the post-test probability of VAP to 0.16, and resulted in $n_T=150$. Estimates from the single centre observational study were used to calculate the sample size. The estimated sensitivity and prevalence in the single centre observational study were $\hat{\lambda}=0.94$ \fw{and} $\hat{\rho}=0.24$, \fw{respectively}, for the most promising biomarker, IL\fw{-}1$\beta$. \kw{If instead} the sample size had been chosen based on a confidence interval for the sensitivity, \fw{using the Wald interval \citep{Zhou2011},} a larger sample size of 196 would have been required.

\subsection{Choosing the sample size \fw{using} assurance}

\begin{figure}
    \centering
    \includegraphics[width=0.45\linewidth]{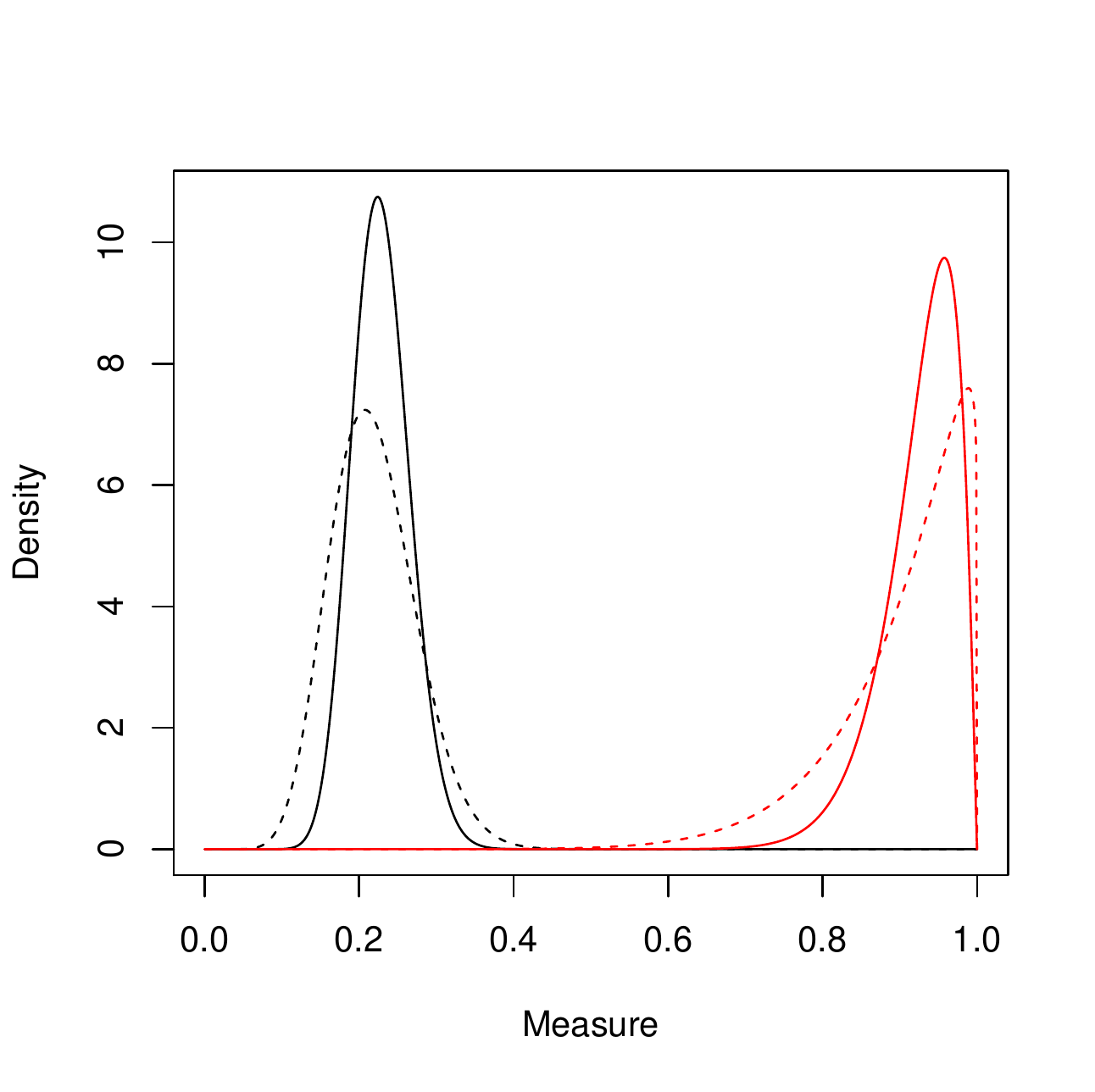}
    \includegraphics[width=0.45\linewidth]{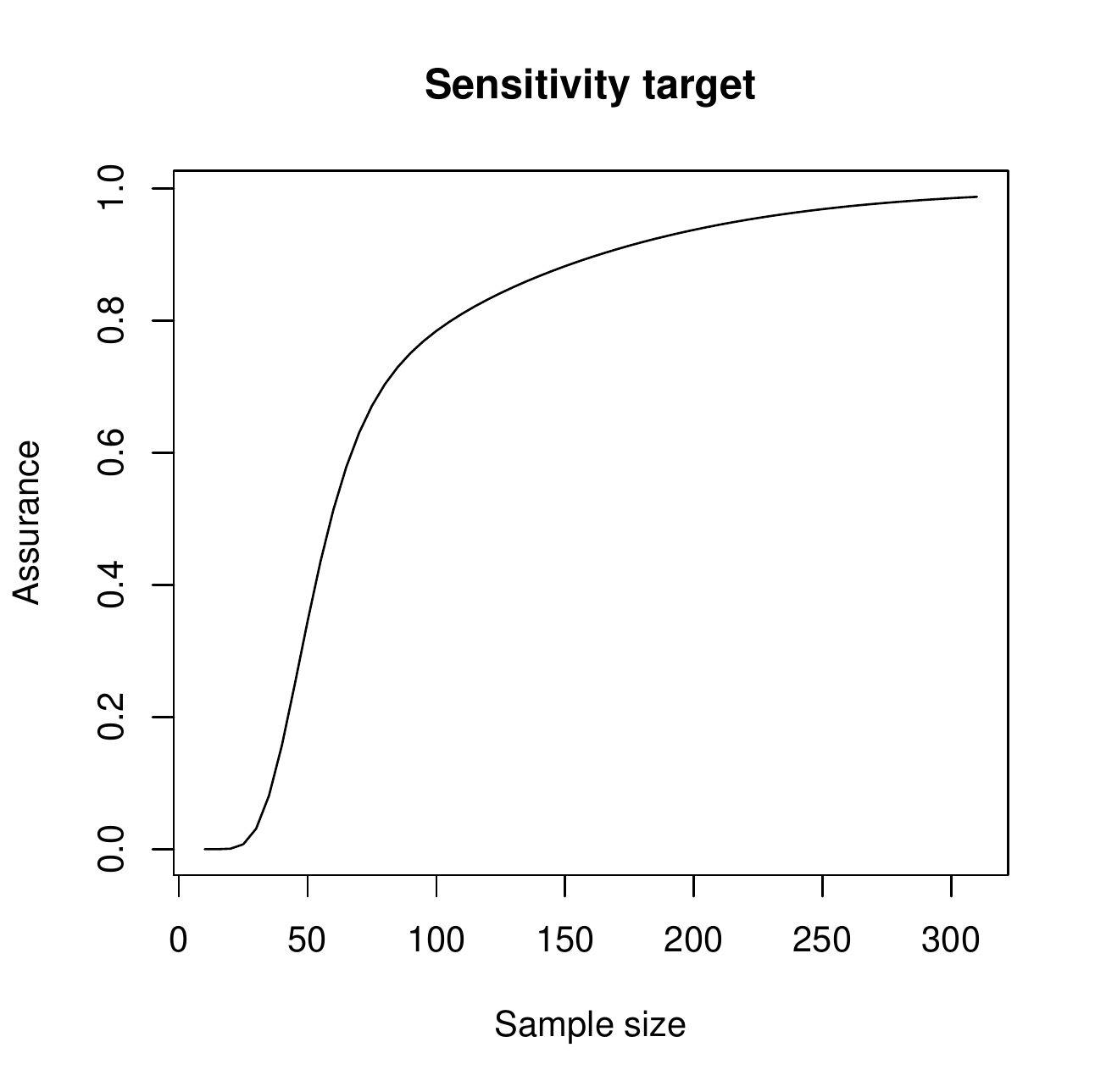}
    \caption{Left:\ The prior distributions for the sensitivity (red) and the prevalence (black) for the biomarker selection study (dashed lines) and the diagnostic accuracy study (solid lines). Right:\ The assurance curve showing the assurance achieved at different sample sizes for the diagnostic accuracy study.}
    \label{priors}
\end{figure}

To use assurance to determine the sample size, we require the prior parameters for the sensitivity, $(a_\lambda^{0},b_\lambda^{0})$, and the prevalence, $(a_\rho^{0},b_\rho^{0})$, before the biomarker selection study. In the initial exploratory study, there were 55 patients, 12 of whom were confirmed by the reference test to have VAP. Assuming exchangeability, a suitable prior for the prevalence is $\rho\sim\textrm{Beta}(12,43)$. The most promising biomarker gave an estimated sensitivity of $0.93$. \fw{Since it was unclear} which biomarker(s) would be used in the final test, it is not reasonable to make an exchangeability assumption for the test results in the two stages\fw{. A more} suitable prior for the sensitivity is more diffuse but with a mean around this value, such as $\lambda\sim\textrm{Beta}(9.9,1.1)$. These priors are \fw{represented by} the dashed lines in Figure \ref{priors}.

In the biomarker selection study, the $2\times2$ contingency table is provided \fw{in} Table \ref{2by2} (ii) \fw{for the most promising biomarker, IL-1$\beta$.}

We assume that these patients are exchangeable with those in the diagnostic accuracy study as they are randomly sampled from the same population. Therefore, the prior distributions for the diagnostic accuracy study are $\lambda\sim\textrm{Beta}(25.9,2.1)$ and $\rho\sim\textrm{Beta}(29,98)$ (see Section \ref{inference}), \fw{illustrated by} solid lines in the left hand side of Figure \ref{priors}.
Suppose we would like to estimate the sensitivity of the test to within 0.16 in a 95\% symmetric probability interval and choose a sample size to give 80\% assurance. Based on the priors above, we use the BAM to obtain a sample size of $n_T^{*} = 106$. This is significantly smaller than the original sample size of $n_T = 150$ (which would give an assurance of 88\%). The full assurance curve for is provided in the right hand side of Figure \ref{priors}. \fw{Note} that the assurance curve has a different shape to a power curve, and is monotonically increasing between 0 and 1.

\subsection{Prior sensitivity} \label{prior_sens}

\fw{To assess the sensitivity of the sample size and assurance to the prior distribution, we use} the approach \fw{outlined} in Section \ref{sensitivity}. \fw{In particular,} we conduct a grid search for both the sensitivity and prevalence priors using a value of $\epsilon=0.00354$ (equivalent to a mean shift in a standard normal random variable of 0.1).

The resulting values of the beta distribution parameters $(a,b)$ are provided \fw{in Section B.3 of the supplementary material for the sensitivity and prevalence}. 
\fw{The corresponding smallest and largest values of the assurance and sample size are provided in Table \ref{sens.result}.}

\begin{table}[ht]
    \centering
    \resizebox{0.4\textwidth}{!} {
    \begin{tabular}{|c|cc|} \hline
Measure & $\min\{A(n_{T})\}$ & $\max\{A(n_{T})\}$  \\ \hline
Sensitivity & 0.73 & 0.86 \\
Prevalence & 0.80 & 0.81 \\ \hline
Measure & $\min\{n_{T}\}$ & $\max\{n_{T}\}$  \\ \hline
Sensitivity & 82  & 130 \\
Prevalence & 104 & 108 \\ \hline
    \end{tabular}
    }
    \caption{The smallest and largest values of the assurance, $A(n_{T})$, and the smallest and largest sample sizes, $n_{T}$, found in the local sensitivity analysis.}
    \label{sens.result}
\end{table}


Changes to the prevalence prior has little effect on the sample size or the assurance at $n_T=106$. \fw{The effect is slightly larger for the sensitivity prior} but, even for the most extreme prior, a sample size of $130$ would be sufficient \fw{(which is considerably less than the sample size of 150 in the study).} 

\subsection{Prior-data conflict} \label{priordataconflict}

\fw{The results from the diagnostic accuracy study with the 150 patients are summarised in Table \ref{2by2} (iii) for the biomarker IL-1$\beta$.}

The resulting posterior distributions for the sensitivity of and prevalence are
$\lambda\sim\textrm{Beta}(76.9,4.1)$ and
$\rho\sim\textrm{Beta}(82,195)$, \fw{respectively}. The \fw{corresponding} 95\% posterior probability interval for the sensitivity is $(0.893,0.986)$, and so we meet the target \fw{of 0.16} on the width of the interval. To assess \fw{possible prior-data conflict, we use the approach detailed in Section \ref{conflict} and compare the observations to the prior predictive distributions.}

\begin{figure}
    \centering
    \includegraphics[height=2.5in]{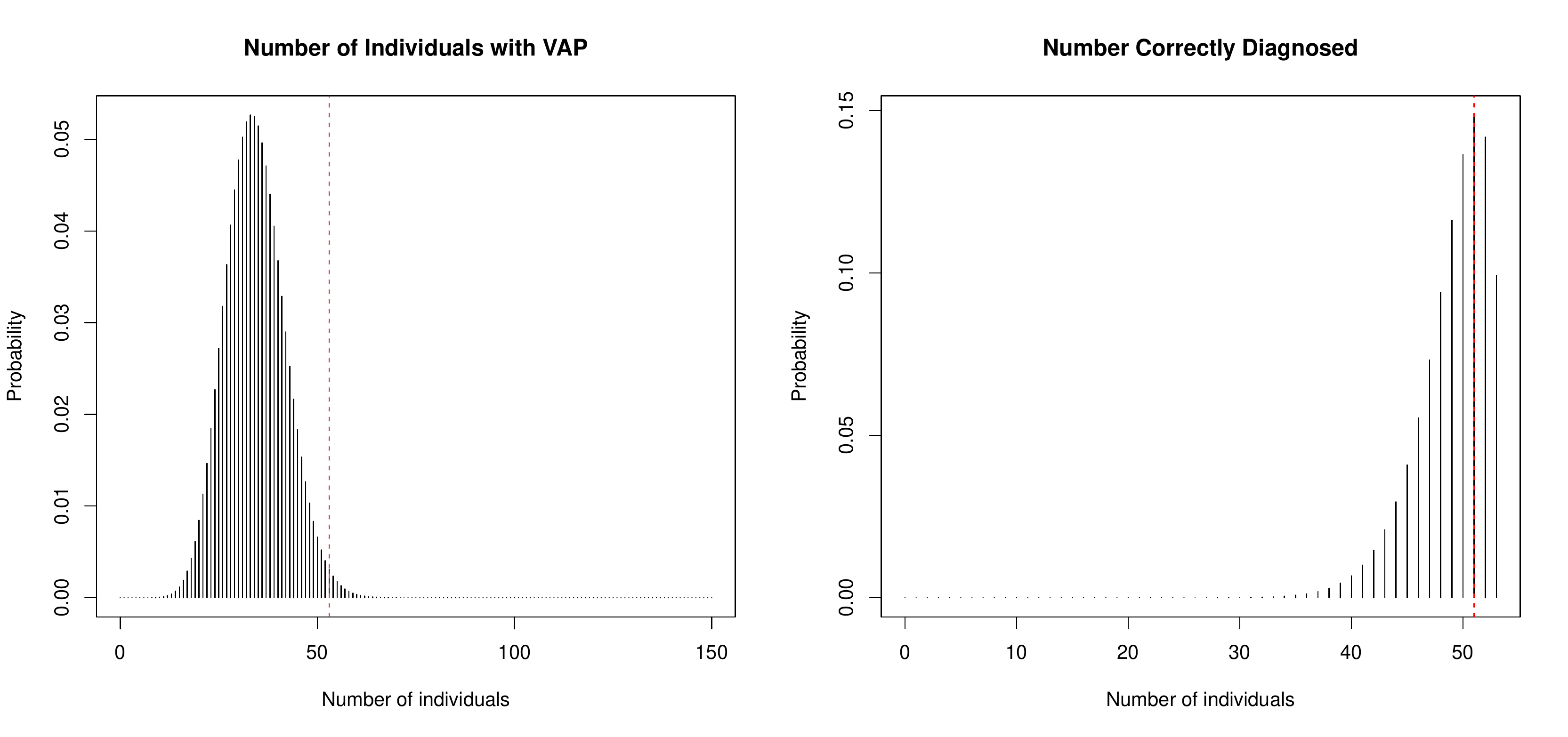}
    \caption{The prior predictive distributions of the number of patients with VAP (left) and the number of VAP patients who test positive (right) together with the observations (red).}
    \label{Conflict}
\end{figure}

The prior predictive distributions of the number of patients with \fw{VAP} (left) and the number of patients with \fw{VAP} who tested positive (right) \fw{are provided} in Figure \ref{Conflict}, \fw{with the observation shown as a red dashed line.} A color version of this figure can be found in the electronic version of the article.

We see the number of patients correctly diagnosed with VAP lies \fw{within} the main body of the \fw{prior} predictive distribution. The \fw{observed} number of patients with VAP lies in the body of the distribution, but is closer to the upper tail, \fw{in the 99th percentile}. The \fw{observed} number of patients correctly diagnosed lies in the 76th percentile. \fw{This provides} some evidence of prior-data conflict for the number of patients with VAP, so we may choose a prior on the prevalence which is not based on the single centre observational study. 

The posterior mean and 95\% posterior probability interval for the prevalence are 0.296 \fw{and} (0.244, 0.351), \fw{respectively}. The same quantities using a flat prior with $a_\rho=b_\rho=1$ are 0.355 \fw{and} (0.281, 0.433),  \fw{respectively}, \fw{which} would not affect the inference on the sensitivity. \fw{However,} if we believe the sub-populations with \fw{VAP} are different between the two stages we would also consider an alternative prior for sensitivity.

\section{Alternative approaches} \label{comparison}


In this section, we compare properties of the \fw{proposed BAM to alternative commonly used methods}. Assume we wish to \fw{obtain the number of individuals with the disease, $n_{T,1}$, required to estimate the sensitivity within a particular degree of accuracy}. The alternative methods are based on a hypothesis test of $H_0:\lambda=\lambda_0$ against the two-tailed alternative $H_1:\lambda\neq\lambda_0$ conducted at a significance level of $\alpha$. \fw{We take the value of $\lambda_0$ to be $\hat{\lambda}$, i.e.\ the maximum likelihood estimate of sensitivity using the analytical validity data.} The sample size \fw{can} be chosen \fw{according to a desired} power of $\beta$ to detect a difference of size $w^{*}$. \fw{As discussed in Section \ref{intro}, there are several possible \kw{approaches}; we consider the following.}

The first is based on a Normal approximation. In this case, to achieve a power of $\beta$ we choose the sample size in the disease group as
$n_{T,1} = \left[\left\{(z_{\alpha/2}+z_{\beta})\sqrt{\hat{\lambda}(1-\hat{\lambda})}\right\}^2\right]/(w^{*}/2)^2$,
where $z_{\cdot}$ is the upper percentile of a standard normal distribution. 
We construct a $100(1-\alpha)$\% confidence interval based on this Normal approximation, known as the \textit{Wald interval} 

The second approach is based on an exact binomial test \fw{to give} the Clopper-Pearson 
(CP) interval. 
The third approach \fw{combines the Normal approximation with an adjustment to the hypothesised value as the centre of the interval to give the Agresti-Coull (AC) interval}. 

\fw{In practice, the standard way of obtaining the required sample size is to use the appropriate sample size formula (if available), or in-built functions within statistical software (e.g.\ the \texttt{binDesign} function from the \texttt{binGroup} R package \citep{Zha18}). However, these often give rise to unreliable sample sizes and, in our investigation, are shown to perform poorly over the range of parameter values considered; see Section E of the supplementary material. We instead rely on simulation. That is}, we choose the smallest sample size $n_{T,1}$ to give the correct proportion of intervals below the desired target width \fw{$w^*$}, based on simulating confidence intervals repeatedly and finding the power empirically. The \textit{total} number of individuals to recruit, $n_T$, \fw{is found by scaling with respect to} the estimated prevalence \fw{$\hat{\rho}$}, i.e.\ $n_T=n_{T,1}/\hat{\rho}$. \fw{The same procedure is used to obtain the number of individuals without the disease, $n_{T,2}$, required to estimate the specificity to a certain degree of accuracy. In this case, $n_T=n_{T,2}/(1 - \hat{\rho}$).}

\subsection{Comparison of sample sizes} \label{priorss}

\begin{figure}[h!]
    \centering
    \includegraphics[width=\linewidth]{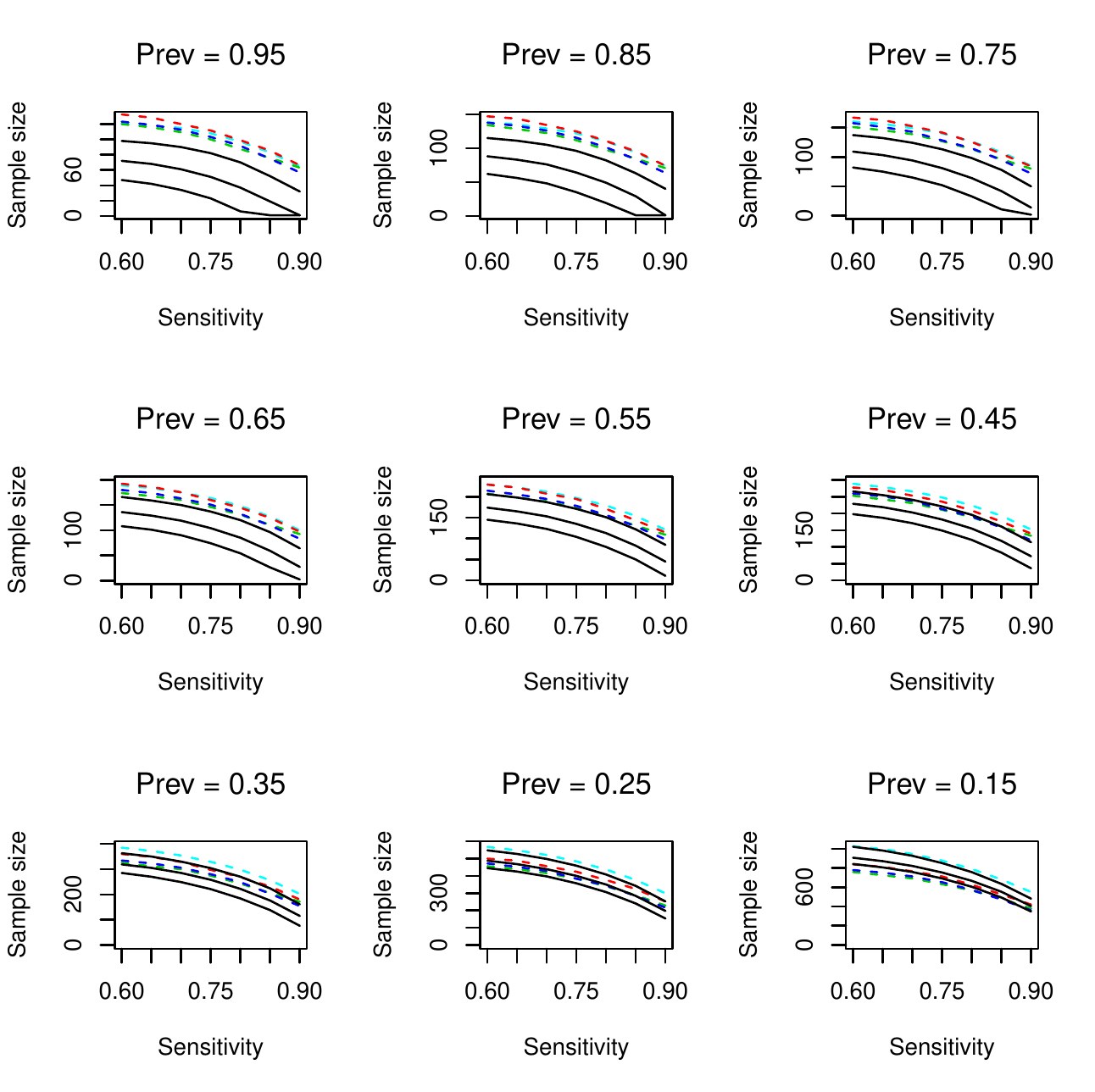}
    \caption{A comparison of the sample sizes required based on power calculations (dashed) using a Wald interval (dark blue), Clopper-Pearson (red), \fw{Agresti}-Coull (green), assurance (black), and assurance based on non-informative analysis priors (light blue). In each plot, there are three black curves relating to prior sample sizes of (from top to bottom) 25, 50 and 75.}
    \label{investigate}
\end{figure}

\fw{In this section}, we compare the sample sizes required for a diagnostic accuracy study \fw{using} the \fw{methods outlined above}. We consider a significance level \fw{of} $\alpha=0.05$, a power/assurance of $\beta=0.8$, and aim to estimate sensitivity to within $0.18$ in a two-sided interval. We vary the sensitivity \fw{over} the range $[0.6,0.9]$ and the prevalence \fw{over} the range $[0.15,0.95]$. For the \fw{proposed} BAM, we consider \fw{three} prior sample sizes \fw{of} $n_T^0=25, 50$ \fw{and} $75$ \fw{to represent} ``small'', ``medium'' and ``large'' analytical validity studies. The \fw{results for all scenarios and methods are illustrated} in Figure \ref{investigate}. 

\fw{Note that} the power calculations are based on the true parameter values. The assurance calculation, \fw{however,} uses beta priors with parameters $(n_T^0\lambda\rho, \; n_T^0\rho[1-\lambda])$ for the sensitivity and $(n_T^0\rho, \; n_T^0[1-\rho])$ for the prevalence. An assurance calculation with non-informative priors for the analysis is also considered. This is based on a design prior from the ``small'' analytical validity study to represent a reasonable ``worst case" scenario. 

\fw{In Figure \ref{investigate}, we observe similar patterns across the frequentist approaches (represented by the coloured lines) for each prevalence.} CP always results in the largest sample size, with Wald and AC giving similar, slightly smaller, sample sizes. In comparison to assurance, the frequentist methods produce larger sample sizes when the prevalence is high. \fw{In some scenarios,} they result in smaller sample sizes. \fw{For example,} when the prior sample size is 25 below a prevalence of 0.5, when the prior sample \fw{size} is 50 below a prevalence of 0.3 and when the \fw{prior} sample size is 75 around a prevalence of 0.2. However, as the sensitivity \fw{increases}, the \fw{required} sample size based on assurance reduces \fw{quicker} than the frequentist approaches, which are known to perform poorly as the sensitivity approaches one.

\fw{Further} details are provided in Section C of the supplementary material, including an assessment of different target interval widths. The message is consistent across the parameter combinations considered:\ assurance for the sensitivity reduces the required sample size in the majority of cases, particularly in moderate to high prevalence populations and when a highly accurate test is required. \rev{High prevalence situations are common in secondary care, where patients have already been triaged (such as in a suspected stroke \citep{shaw2021purines}), or in cancer pathways by the time an invasive test, such as a biopsy, is used. When the BAM is applied to even lower prevalences of 0.1, 0.05 and 0.01, the sample sizes required for a sensitivity of 0.9, and based on a medium analytic validity study, are 681, 1643 and 2770, respectively. Such low prevalences may be the case in large-scale geographic prevalence surveys, for example.} 

\subsection{Comparison of interval widths \label{intervalwidths}}

A smaller sample size will not be useful if the \fw{corresponding} interval estimates are very wide. Therefore, we conduct a simulation study, \fw{outlined below,} to assess the width of the intervals resulting from each approach. 

\begin{figure}[h!]
    \centering
    \includegraphics[width=0.45\linewidth]{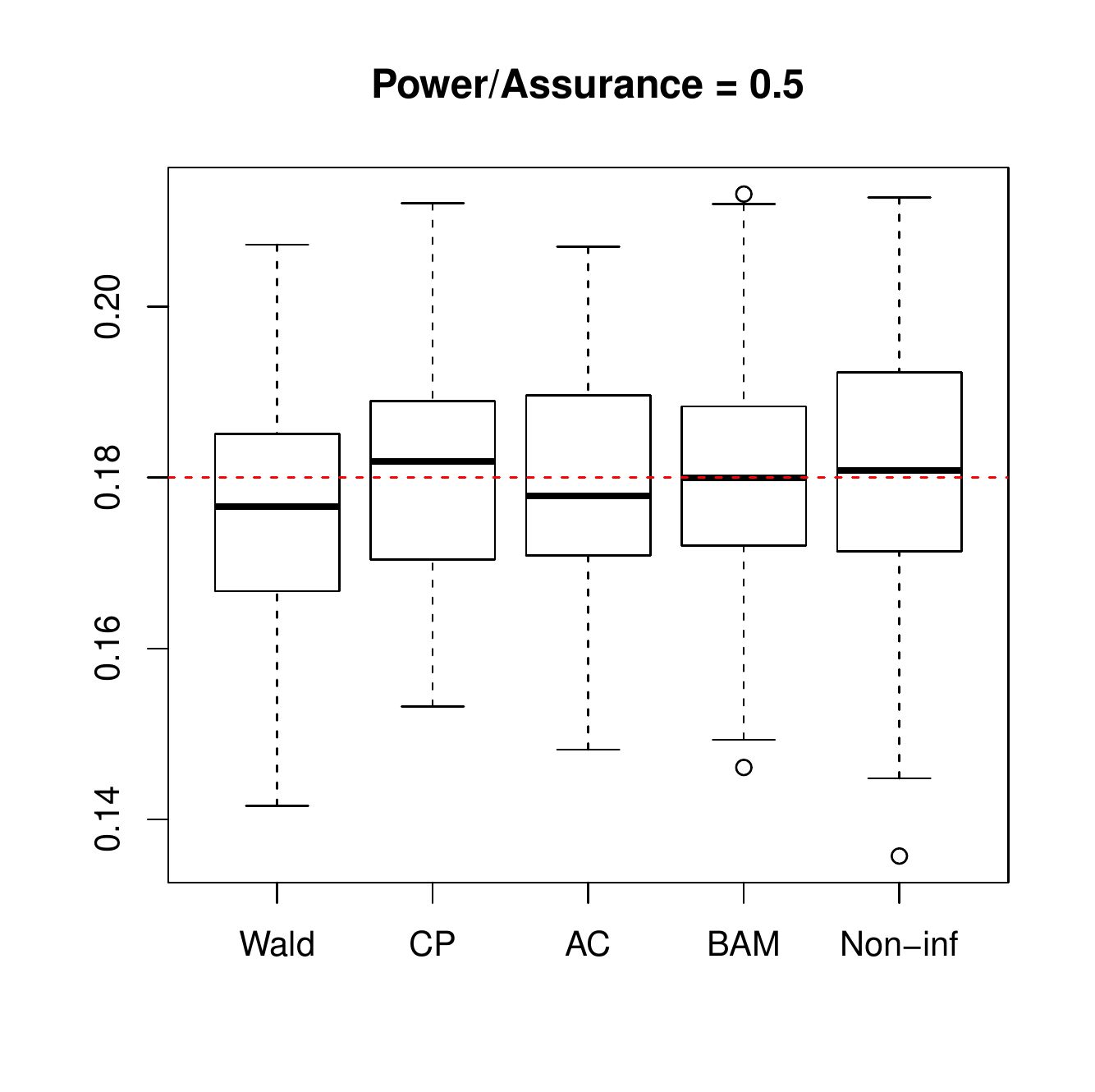}
    \includegraphics[width=0.45\linewidth]{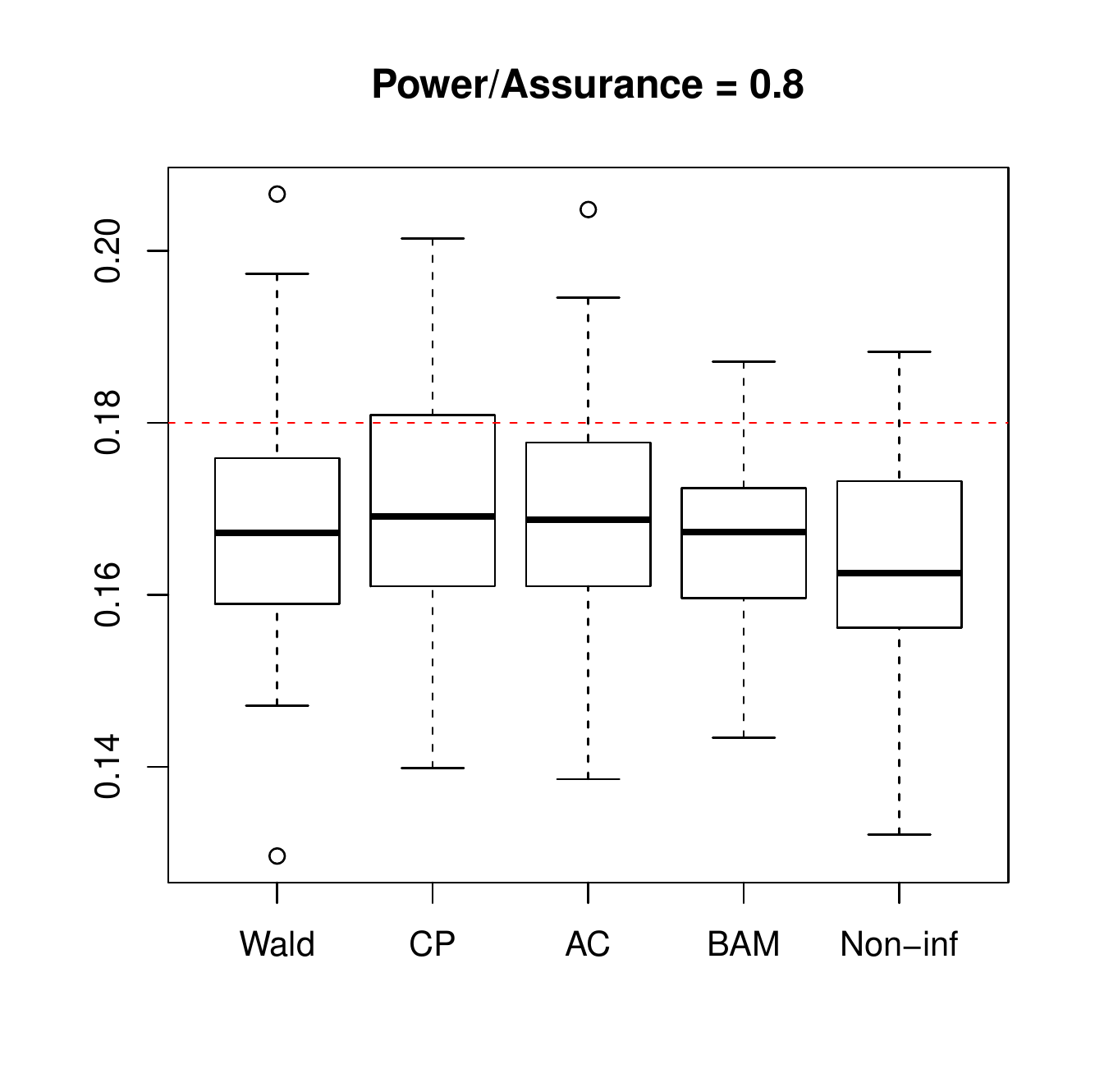}
    \caption{The width of 95\% confidence or posterior probability intervals based on $100$ simulations for the Wald interval (Wald), Clopper-Pearson \fw{(CP)}, Agresti-Coull \fw{(AC)}, Assurance \fw{(Bayes)} and Assurance using a non-informative analysis prior \fw{(Non-inf)}. The power/assurance used to choose the sample size was 0.5 (left) and 0.8 (right). The horizontal line is at the desired width of $w^{*}=0.18.$.}
    \label{simulate}
\end{figure}

First, we sample values of the sensitivity and prevalence from uniform distributions. These \fw{are used} to sample analytical validity results, $n_{1,1}^0$ and $n_{T,1}^0$, from their respective binomial distributions based on a ``medium'' total sample size of $n_T=50$. From these data, we find estimates of the sensitivity and prevalence for the power calculations and set the prior distributions for the assurance calculations. We then find the required sample size for each method. We sample the results of the diagnostic accuracy study, $n_{1,1}^1$ and $n_{T,1}^1$, from their respective binomial distributions and use these to calculate $100(1-\alpha)$\% intervals for the sensitivity. Finally, we calculate the width of the interval\fw{s}. By repeating this process $100$ times, we consider the distributions of widths of the intervals, \fw{which are shown} in Figure \ref{simulate} for a power/assurance, $\beta$, of 0.5 (left) and 0.8 (right). In all cases, $\alpha=0.05$ and $w^{*}=0.18$. 

\fw{For} both $\beta=0.5$ and $\beta=0.8$, the approaches produce intervals with a similar distribution of widths. When $\beta=0.5$, the median width of each approach \fw{lies} approximately at the target width. When $\beta=0.8$, the target width is around, or slightly above, the upper quartile for each method. Thus, the different sample sizes observed in the previous section do not come at the expense of less precision in inference.


The simulations were repeated with interval widths of $w^{*}=0.22$ and $w^{*}=0.14$. The \fw{corresponding} results \fw{are} provided in Section D of the supplementary material. The main conclusions remain:\ for a power/assurance of 0.5, all of the distributions \fw{are} approximately centred on the target width, and for a power/assurance of 0.8, each approach produces intervals which include the target width in the upper 25\% of its empirical distribution. 

\rev{In addition, we have investigated the properties of the BAM when assuring both sensitivity and specificity together, in terms of the sample size required and the resulting interval widths. This is provided in Section F of the supplementary material.}

\section{Discussion}
\label{discussion}

\fw{In this paper, we have proposed the novel BAM to determine sample sizes for diagnostic accuracy studies.} Bayesian assurance fulfils a similar role \fw{to} power and, as we have shown, can offer benefits when suitable prior information is available. \fw{In particular, representing uncertainty in unknown test characteristics using prior distributions, and utilising information from different stages of the development pathway,} allows for a wider range of evidence to be \fw{seamlessly incorporated} in the design and analysis of a diagnostic accuracy study. \fw{Consequently, we have shown that this has the potential to reduce the sample size, thus increasing efficiency in evidence development.}


\fw{If no prior information} is available, \fw{or accessible, from} earlier stages of development, expert elicitation can be used to form the \fw{necessary} prior distributions.
Elicited distributions can include \fw{opinions} from multiple experts, or be combined with data from other sources \citep{williams2020comparison}. 
The larger the prior sample size, the more informative the prior distribution will be \fw{which, as shown in Figure \ref{investigate},} typically corresponds to a smaller sample size in the \fw{diagnostic accuracy} study. 
If it is not appropriate to use an informative prior for the analysis (e.g.\ to mitigate researcher bias), a sceptical or flat prior can be used instead. \fw{The BAM has the flexibility to} allow for distinct prior distributions in the design and analysis stages, \fw{as illustrated in Section B of the supplementary material.}

\fw{The proposed BAM} can be used regardless of \fw{whether} the final analysis \fw{is} frequentist or Bayesian. Some assurance calculations may not result in closed form solutions (e.g.\ if a Bayesian analysis uses a non-conjugate analysis prior), in which case, simulation and numerical methods are required. 
Thus, calculating assurance can be challenging and, unlike power, is not available in standard software packages. To increase accessibility of the BAM, R code is provided and an R Shiny application is currently under development. 

This work focuses on assuring sensitivity and specificity as measures of diagnostic accuracy. We have also shown how the BAM can be used to assure sensitivity and specificity jointly, for which no existing approaches are available, to our knowledge. The assurance calculations can be modified to obtain sample sizes for other quantities, such as positive and negative predictive values \kw{or the area under the curve}. Moreover, the assurance calculations could be \fw{extended} to allow for multiple categorical results, or results in the form of continuous measures, which forms an area of further work. \fw{In this paper, we considered the evaluation of a single diagnostic test, but further work could explore how the proposed method extends to multiple tests.} 

\rev{To reflect standard practice in diagnostic accuracy studies, we have inherently assumed that the sampling plan will be produced prior to the study, carried out accordingly and then the data analysed at the end of the study. Future work could extend the approach so that it can be applied sequentially, participant-by-participant (or in blocks), to monitor the width of the posterior interval until the desired value is attained, at which point the study would terminate. This would reduce the sample size required. However, it would require a change in the way that diagnostic accuracy studies are routinely implemented.}


\section*{Acknowledgments}
The authors thank Prof.\ James Wason and Prof.\ John Simpson for useful discussions. AJA, CJW, BCL are supported by the NIHR Newcastle In Vitro Diagnostics Co-operative. The funders had no role in the preparation or decision to publish this manuscript. The views are those of the authors and not necessarily those of the NHS, NIHR, or the Department of Health and Social Care. 
The VAPrapid trial was supported by the Department of Health and Wellcome Trust via the Health Innovation Challenge Fund.

\section*{Supporting information}

Additional supporting information may be found online in the Supporting Information section at the end of this article.

R code used to implement the method is also available here. Data sharing is not applicable to this article as no new data were created or analyzed in this study.







\appendix

\section{Derivation of the equivalence of the beta-binomial and multinomial-Dirichlet approaches}

Based on the multinomial-Dirichlet prior structure, the sensitivity and specificity are independent in the prior. To see this, we note that the parameter vector $\bm\gamma$ can be re-ordered so that $\bm{\tilde{\gamma}}=(\gamma_{1,1},\gamma_{2,1},\gamma_{1,2},\gamma_{2,2})^{'}$, which has a Dirichlet prior with re-ordered parameter vector $\bm{\tilde{\alpha}}$. Then, by the properties of neutrality and aggregation of the Dirichlet distribution, we see that $(\gamma_{1,1}/[\gamma_{1,1}+\gamma_{2,1}],\gamma_{2,1}/[\gamma_{1,1}+\gamma_{2,1}])$ and $(\gamma_{1,2}/[\gamma_{1,2}+\gamma_{2,2}],\gamma_{2,2}/[\gamma_{1,2}+\gamma_{2,2}])$  are mutually independent. Hence $\gamma_{1,1}/[\gamma_{1,1}+\gamma_{2,1}]$ and $\gamma_{2,2}/[\gamma_{1,2}+\gamma_{2,2}]$ are independent. Now, since $\bm\gamma$ has a Dirichlet prior distribution, this means that $\gamma_{i,j}\gamma_{\Sigma}\sim\textrm{Gamma}(\alpha_{i,j},1),$
where $\gamma_{\Sigma}=\sum_{i=1}^2\sum_{j=1}^2\gamma_{i,j}$. We can re-express the sensitivity as
$\gamma_{1,1}/(\gamma_{1,1}+\gamma_{2,1}) = \gamma_{1,1}\gamma_{\Sigma}/(\gamma_{1,1}\gamma_{\Sigma}+\gamma_{2,1}\gamma_{\Sigma})$
and, by the properties of the gamma distribution, we have
$ \lambda \sim \textrm{Beta}(\alpha_{1,1},\alpha_{2,1})$.
By similar reasoning, $\theta \sim \textrm{Beta}(\alpha_{2,2},\alpha_{1,2})$ \fw{for} specificity. 

When the number of individuals in each cell of the contingency table is observed, the posterior distributions are
\begin{eqnarray*}
\bm\gamma\mid\bm n & \sim & \textrm{Dir}(\alpha_{1,1}+n_{1,1},\alpha_{1,2}+n_{1,2},\alpha_{2,1}+n_{2,1},\alpha_{2,2}+n_{2,2}), \\
\lambda\mid\bm n & \sim & \textrm{Beta}(\alpha_{1,1}+n_{1,1},\alpha_{2,1}+n_{2,1}), \\
\theta\mid\bm n & \sim & \textrm{Beta}(\alpha_{2,2}+n_{2,2},\alpha_{1,2}+n_{1,2}).
\end{eqnarray*}

Setting $\alpha_{1,1}=a_\lambda$, $\alpha_{1,2}=b_\theta$, $\alpha_{2,1}=b_\lambda$ and $\alpha_{2,2}=a_\theta$, we see that the two approaches are equivalent for the sensitivity and specificity.

\bibliographystyle{plainnat}
\bibliography{refs}%

\end{document}